\documentstyle[preprint,epsfig,aps]{revtex}
\begin{document}
\title{High Energy Cosmic Rays from Neutrinos} 
\author{ J.J.~Blanco-Pillado$^{1,2}$, R.A.~V\'azquez$^2$, and E.~Zas$^2$}
\address{$^1$ Institute of Cosmology \\
Department of Physics and Astronomy, Tufts University\\
Medford, MA 02155 \\
$^2$ Departamento de F\'\i sica de Part\'\i culas, Universidad
de Santiago\\
E-15706 Santiago de Compostela, Spain\\}
\maketitle
\begin{abstract}
We address a class of models in which neutrinos, having a small mass, 
originate the highest energy cosmic rays interacting with the 
relic cosmic neutrino background. Assuming lepton number symmetry and 
an enhanced neutrino density in arbitrary size clusters (halos), 
we make an analytical calculation of the required neutrino fluxes. 
We show that the parameter space for these models 
is heavily constrained by horizontal air shower searches. Marginal room is 
left for models with exceptionally flat neutrino spectral indices, 
neutrino masses in the $0.1~$eV range and supercluster scale halos of 
order 50~Mpc size. Our constraints do not apply to models with 
lepton number asymmetry. 
\end{abstract}
\pacs{96.40.Tv,96.40.Pq,98.70.Sa,98.80.Cq}

The detection of ultra high energy cosmic rays (UHECR) above the 
Greisen-Zatsepin-Kuzmin (GZK) cut off \cite{GZK} has stirred the research 
activity in cosmic acceleration mechanisms. Above the GZK cutoff protons 
rapidly loose energy through photoproduction in the cosmic microwave 
background and therefore sources must be relatively nearby \cite{Stecker}. 
The 
known objects in our ``extragalactic neighborhood'', within few tens of 
Megaparsecs, have difficulties to accommodate stochastic acceleration 
mechanisms (most commonly invoked) due to dimensional arguments. Moreover such 
UHECR deviate little in the magnetic fields encountered over this length scale 
and no obvious astrophysical candidates are seen in the arrival direction 
of the few 
detected events. Several possibilities have been considered to explain these 
events \cite{Sigl,Boldt} 
and in particular that cosmic ray production arises through ultra 
high energy (UHE) neutrino interactions with the cosmic neutrino background 
\cite{Weiler,Fargion}. 

UHE neutrinos could come from cosmic distances and interact with the relic 
neutrinos in our halo. The final stable products of these interactions would 
be gamma rays and protons (besides secondary neutrinos) which would constitute 
the high energy end (above $\sim 10^{19}~$eV) of the cosmic ray spectrum. The 
idea is attractive because it avoids the constraint that source candidates 
must be at distances below $\sim 50~$Mpc, but it requires 
large fluxes of very high energy neutrinos (above $\sim 10^{21}~$eV) 
\cite{Yoshida,Waxman} 
without getting into the details concerning the UHE neutrino 
production mechanism. Models involving annihilation of 
topological defects \cite{Sigl,BerezinskyTD} and heavy relic 
decays \cite{GelminiTD} could for instance produce these 
neutrinos rather naturally. Bounds 
for some of these models have already been discussed in the 
literature based on neutrino and photon flux measurements 
\cite{BerezinskyTD,protstan,Blanco_1,berezinsky2}. 

Only the resonance peak in the $Z^0$ 
production interactions with the cosmic neutrino background can provide any 
significant secondary particle flux. In the massless neutrino case an energy 
$E_\nu^{res} \sim 10^{16} $~GeV is required to produce $Z^0$'s at resonance 
since relic neutrinos have energies $\sim 2$~K $\simeq 1.7~10^{-4} $ eV.  
This possibility would imply either neutrino fluxes exceeding current 
limits from Horizontal Air Showers, as will be shown below, or  
unnaturally fine tuned neutrino energy spectra. 
If neutrinos are massive, a possibility that is becoming 
increasingly more realistic in the light of recent results by Superkamiokande 
\cite{oscilations}, the necessary neutrino beam energy to 
produce $Z^0$'s at resonance in the interactions with the relic neutrinos 
becomes inversely proportional to the neutrino mass. Moreover background 
neutrinos would tend to accumulate in an extended halo as pointed out in Ref. 
\cite{Weiler,Fargion,Weiler_1,Roulet} increasing their local density with 
respect to the cosmological value and the probability of nearby interactions.

The idea has already been discussed by several authors. Using an incoming 
neutrino flux of spectral index 2, Waxman has discussed the models from 
general energy density arguments, using limits on local neutrino density 
because of Pauli exclusion, to conclude that a new class of models would have 
to be invoked to accelerate the neutrinos themselves and that the energy 
required is comparable to the total photon luminosity of the Universe 
\cite{Waxman}. The calculation depends on the spectral index and the energy 
cutoffs of the assumed neutrino spectrum. Yoshida {\sl et al.} have recently 
computed the particle spectra for several case studies after detailed 
propagation of all secondary products in the extragalactic magnetic fields and 
through the cosmic microwave background, and assuming clustering in 
supergalactic scales. These cases support the possibility that the produced 
UHECR, neutrinos and gamma rays are compatible with neutrino observations and 
bounds \cite{Yoshida}. 

In this paper we further discuss this idea analyzing the model dependence on 
the assumed neutrino spectral index. We establish the neutrino fluxes firstly 
using energetic considerations similar to those of Ref.\cite{Waxman} and then 
analytically calculating the proton and photon secondary spectra. If large 
fluxes of high energy neutrinos exist, they should have been detected. Indeed 
by imposing that the neutrinos produce the observed UHECR one expects a much 
larger flux of neutrinos as pointed out in Ref.\cite{Yoshida}, since the 
probability for interacting in the neutrino halo is small. In a 
phenomenological approach we leave the neutrino spectral index and the local 
neutrino density enhancement in the halo as free parameters. We will show that 
for a large region in the two dimensional parameter space, the required 
neutrino flux is heavily constrained by existing data on horizontal showers.

Assuming neutrinos are massive (of order 1~eV) we consider a neutrino of 
energy $E_\nu$. This neutrino could interact with an antineutrino from the 
cosmic background, with a center of mass squared energy $s = 2 m_\nu E_\nu $. 
The cross section for this process is maximal near the $Z^0$ resonance, of 
width $\Gamma_Z$, which occurs for neutrino energies of $E_\nu \sim 4 \; 
10^{12} (1 \; \mbox{eV}/m_\nu) $ GeV. In hadronic decays the $Z^0$ produces 
high energy particles, mostly pions, which further decay so that only high 
energy photons and protons (and neutrinos) would eventually reach the Earth. 
The final particle spectra are given by a convolution of the quark 
fragmentation functions and the in flight decays of all the intermediate 
particles. These spectra have to be propagated in the photon cosmic background,
IR background, galactic fields, etc. which would alter the arrival fluxes of 
high energy photons and protons. As shown by Yoshida et al. \cite{Yoshida}, the
final particle spectra agree well with observations and can explain the UHECR
spectrum for a wide range of spectral indices, $\gamma \sim 2$, assumed in the 
original neutrino flux. 

The survival probability of a UHE neutrino in the relic neutrino background is 
in general given by ${\cal P_S}(E_{\nu})=e^{- \tau_{\nu}}$, with $\tau_{\nu}$ 
being the opacity. Considering only the resonant $Z^0$ production cross 
section $\sigma_{\nu \bar \nu}$ \cite{foot} in a matter dominated Universe 
the opacity can be well approximated by the redshift integral:
\begin{equation}
\tau_{\nu}  \left(E_{\nu 0} \right) \simeq {c \over H_0} ~n_{ \bar \nu 0}~
\int _0 ^{z_{max}} \!\!\!\!\! 
{dz (1+z) ^2~
     \sigma _{\nu { \bar \nu}} \left[2 m_{ \bar \nu} E_{\nu 0} (1+z) \right]
      \over \sqrt{\Omega_{M 0} (1+z)^3 + 
            \left[1-\Omega_{M 0} -\Omega_{\Lambda 0} \right] (1+z)^2 
            + \Omega_{\Lambda 0}}} 
\label{opacity}
\end{equation}
Here $H$ is the Hubble constant, $E_{\nu}$ is the interacting 
neutrino energy and the subscript $0$ is used to indicate 
the present value of a redshift varying quantity. 
$\Omega_M$ and $\Omega_{\Lambda}$ are respectively 
the matter density and the cosmological constant 
terms in the Friedmann equation expressed in dimensionless units. 
The energy integral over the relic neutrino spectral density has been 
eliminated in the assumption the neutrinos are non relativistic 
so that $n_{\bar \nu}$ is the relic antineutrino number density and 
the argument of the interaction cross section is the redshift 
varying center of mass energy of the collision.  

If one integrates this expression to galaxy formation era
$z_{max} \sim 5$ assuming no clustering of the relic neutrinos, 
the uncertainty in the numerical value of the opacity is mainly 
dominated by the lack of precise knowledge of $\Omega_{M 0}$ and $H_0$. 
For incoming neutrinos having the 
appropriate energy to interact resonantly, the opacity obtained 
ranges from $\sim 0.05$ 
to $\sim 0.3$ for cosmological scenarios with $ 0.1 < \Omega_{M 0} < 1$ and 
cosmological constant parameter in the range 
$ 0 < \Omega_{\Lambda 0} < 0.7$~\cite{opacity}. 
Although models in which the UHECR are produced by UHE neutrinos can 
require energy densities comparable to the luminosity of the Universe 
\cite{Waxman}, they would not 
necessarily have strong observable effects provided the opacity is low. 
If the opacities were larger, as one could expect for a mechanism 
generating the neutrinos at higher redshifts, there could be other 
observable consequences such as low energy photons above current 
experimental limits. 
Whatever the origin of the interacting neutrinos, if one requires a 
neutrino flux well exceeding that of other particles such as electrons, photons, protons and neutrons, models can be found which are consistent 
with the low energy photon flux bound as shown by the examples in 
Ref.~\cite{Yoshida}. 

In order to explain the UHECR spectrum nearby the Earth the production rate 
within the absorption distance of the cosmic rays in the CMB ($\sim 50$~Mpc) is
fixed by data. If the local relic neutrino density is known this normalizes the
neutrino flux. We leave a local density enhancement factor, $10^{\xi}$, to 
account for possible clustering effects and assume a halo radius $D$, otherwise
the probability of interaction for a neutrino is very small and the neutrino 
flux needed to produce the cosmic rays must be enormous and in conflict with 
both energy considerations and experimental neutrino bounds. The survival 
probability for the incoming neutrino flux is given by a local opacity factor 
$\tau_{D \nu}$ integrating Eq.(~\ref{opacity}) to the halo limit taken to be 
$D$. This probability has a large resonance peak at 
$E_{\nu}^{res}=M_Z^2/(2m_{\nu})$ of width $\delta E_{\nu}=E_{\nu} \Gamma_Z /
M_Z$ where $M_Z$ and $\Gamma_Z$ are the $Z$ mass and width respectively. As 
long as $D$ is below 50~Mpc the upper $z$ limit is 
small $z<0.01$ and the opacity at the resonant energy can be well 
approximated by the (''static'') expression: 
\begin{equation}
\tau_{D \nu} \left(E_{ \nu 0} \right) \simeq  
D~\sigma _{\nu { \bar \nu}} \left( 2 m_{ \bar \nu} E_{ \nu 0} \right) 
~n_{ \bar \nu 0} \simeq  1.3~ 10^{-5}~10^\xi~
{D \over 1~\mbox{Mpc}}.
\label{localopacity}
\end{equation}
The probability of interacting locally in the halo is given by 
${\cal P}_I (E_{\nu})=1-e^{-\tau_{D \nu}} \simeq \tau_{D \nu}$. 
As long as its size is not extremely large, $D \lesssim 50 $ Mpc, 
the interaction probability, ${\cal P}_I$, is small except for very 
large density enhancement factors.  

Taking the local neutrino flux entering the halo region to be 
$\phi(E_{\nu})$, the 
injection energy through resonant $Z^0$ production is simply given by: 
\begin{eqnarray}
{\cal E}_\nu &= & \int dE_{\nu}~E_{\nu}~{\cal P}_I(E_{\nu})~\phi(E_{\nu}) 
\nonumber\\
&\simeq & 1.3~[E_{\nu}^{res}]^2~{\cal P}_I(E_{\nu}^{res})~\phi(E_{\nu}^{res})
~{ \Gamma_Z \over M_Z},
\label{eneutrino}
\end{eqnarray}
where the last expression corresponds to the common approximation used for 
integrating the resonant cross section and the factor makes the expression 
numerically exact for a neutrino spectral index of 2. Following Waxman, 
Eq.(\ref{eneutrino}) can be equated to the produced energy flux of cosmic rays 
to obtain $\phi(E_{\nu}^{res})$:
\begin{equation}
{\cal E}_p = \int_{E_{min}}^{E_{max}} dE  \; E \; \phi_p(E),
\label{eproton}
\end{equation}
where $\phi_p(E)$ is the higher energy cosmic ray flux tail assumed to be
due to this mechanism. For a neutrino spectral index $\gamma$, the flux is:  
\begin{equation}
\phi_\nu(E) = {{\cal E}_p \over 1.3~[E_{\nu}^{res}]^2~{\cal
P}_I(E_{\nu}^{res})~{ \Gamma_Z / M_Z}}~
\left[ {E_{\nu} \over E_{\nu}^{res}} \right]^{-\gamma}.
\label{phinu}
\end{equation}

\begin{figure}
\begin{center}
\mbox{\epsfig{file=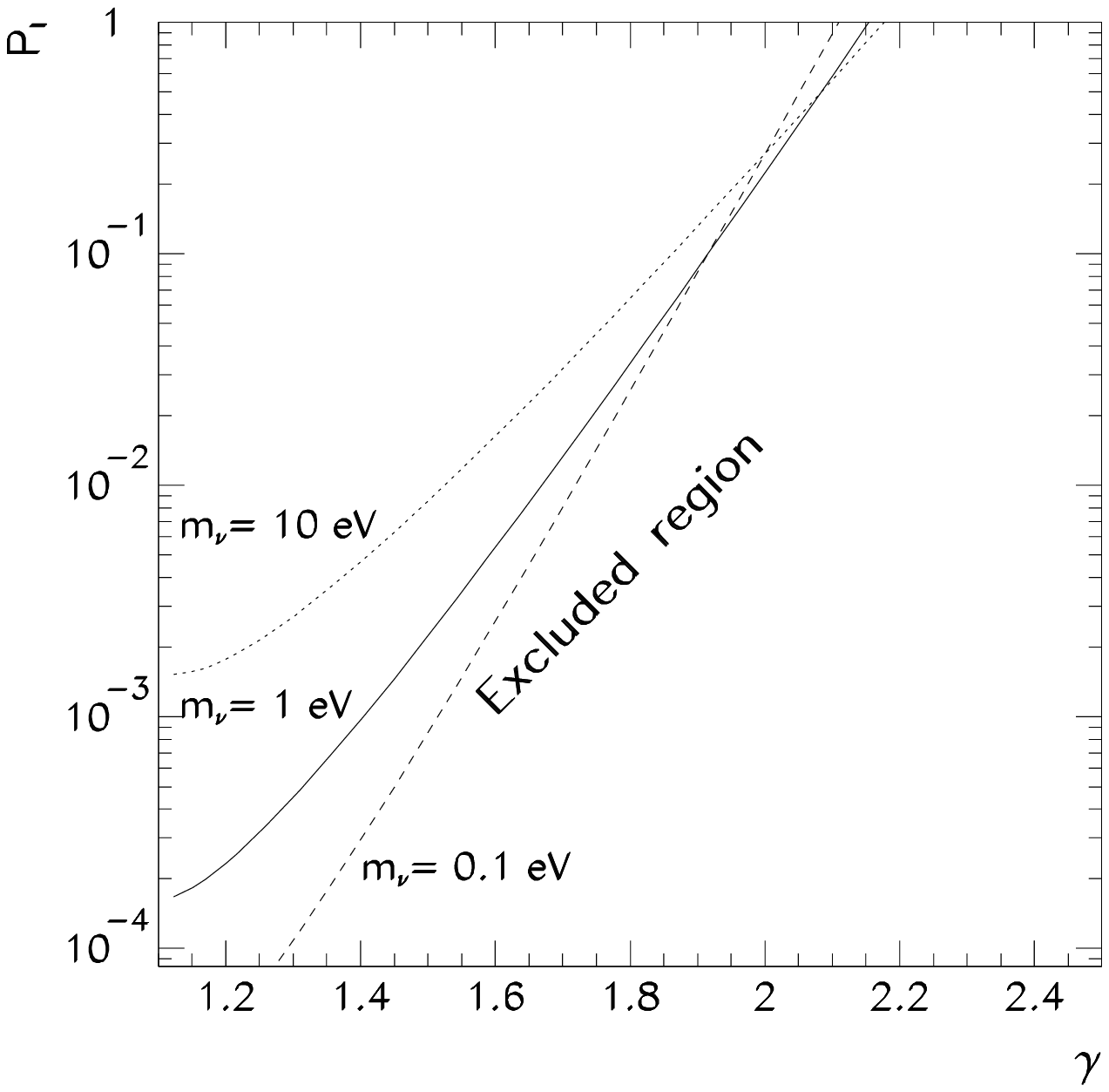,width=0.3\textwidth}}
\end{center}
\caption{Excluded region in the ${\cal P}_I,\gamma$ parameter space for 
three different values of the neutrino mass as marked. }
\label{fig1}
\end{figure}

The important point is that $\phi(E_{\nu}^{res})$ is inversely proportional to 
the interaction probability at the resonance peak ${\cal P}_I(E_{\nu}^{res})$ 
and to its width $\delta E_{\nu}$. One should expect extrapolations of this 
flux with a fairly constant spectral index $\gamma$ both below and above the 
resonant energy because experience in astrophysical fluxes and theoretical 
considerations are very strong in supporting a neutrino flux spanning a few 
decades in energy. High energy fluxes from low interacting particles are 
severely constrained by existing experiments \cite{Blanco_1}. For the range of 
energies considered here the strongest limit is given by the Fly's Eye group 
\cite{FlysEye}. The non observation of horizontal air showers allows to put a 
limit on the integrated flux of any low interacting particle. Provided the 
neutrino flux can be extrapolated to the effective energy threshold for the 
Fly's Eye bound, $E_F \sim 10^8 $ GeV:
\begin{equation}
\Phi_\nu(E_F) = \int_{E_F}^\infty dE \phi_\nu(E) 
\left[ 1 - {\cal P}_I(E_{\nu}) \right]  \le \Phi_0.
\label{limit}
\end{equation}
Fixing the neutrino mass, assuming $\gamma > 1$, choosing a conservative 
(high) value of $E_{min}= 5 \; 10^{19}$~eV, and $2.5$ for the proton spectral
index in Eq.(~\ref{eproton}), we can constrain the region of allowed values of 
$\gamma$ and ${\cal P}_I$, using Eq.(~\ref{phinu}) for $\phi_{\nu}$. This is 
shown in fig.~\ref{fig1} for $m_\nu= 0.1, 1$, and 10 eV. The figure shows that 
there is a critical spectral index $\gamma = 2.15$ above which the model is 
ruled out for all masses in the $0.1-10$~eV range. That is if $\gamma > 2.15$ 
horizontal showers should have been observed even in the event that all 
neutrinos in the resonant energy range were converted to UHECR. 
If $\gamma \simeq 1.2$ however, a very low 
(depending on the neutrino mass) conversion probability could be allowed by 
data. 

\begin{figure}
\begin{center}
\mbox{\epsfig{file=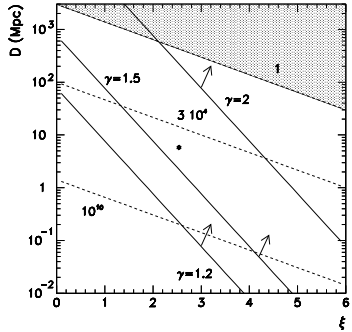,width=0.3\textwidth}}
\end{center}
\caption{Full lines mark the excluded regions by HS data (below the curves) 
in the $D$~(Mpc), $\xi$ parameter space for different values $\gamma$ assuming 
$m_\nu$= 1 eV. The neutrino flux has been normalized to the total energy 
in the UHECR (see text). The dashed lines represent the maximum number of 
halos provided 
the total number of relic neutrinos is fixed. The star marks the position of 
model in Ref.[9].} 
\label{fig2}
\end{figure}

Using Eq.(~\ref{localopacity}) to substitute the probability 
${\cal P}_I(E_{\nu})$ into the experimental limit expression in 
Eq.~(\ref{limit}) we get a region of allowed parameter space $\xi,D$ for any 
given value of $m_\nu$ and $\gamma$. This is shown in fig.~\ref{fig2} where 
the limits are given as the continuous lines for different values of the 
neutrino spectral index.

Further restrictions apply in this parameter plot. The maximum density is 
constrained by the Fermi distribution to be \cite{Waxman,Peebles}:
\begin{equation}
n_{\nu 0} \le 1.5 \; 10^3 \left( \frac{m_\nu}{1 \; \mbox{eV}} 
\right)^3 \left( \frac{v}{220 \; \mbox{km/s}} \right)^3 \mbox{cm}^{-3},
\end{equation}
where $v$ is the characteristic neutrino velocity in the halo.
However, given the strong dependence on the neutrino mass and an unknown 
velocity we take it as a free parameter. 

\begin{figure}
\begin{center}
\mbox{\epsfig{file=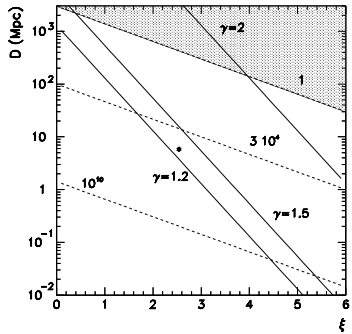,width=0.3\textwidth}}
\end{center}
\caption{Same as fig.2, with neutrinos normalized to the UHECR flux 
(see text).}
\label{fig3}
\end{figure}

In addition, if the total number of background neutrinos in the Universe is 
fixed, the density enhancement factors in the halo, their sizes and the 
maximum total number of halos are related. For constant density halos, 
assuming that no neutrinos are outside them, the number of neutrino clusters, 
in a Hubble radius, of a given size and a given enhancement factor is simply 
$N_c= 10^{-\xi} (R_H/D)^3$, where $R_H$ is the Hubble radius. 
One can now 
easily see that the maximum number of halos can be read in a slant coordinate 
shown as dashed lines in Figs.~\ref{fig2},\ref{fig3}. The three lines correspond to a 
single halo, $3\;10^4$ and $10^{10}$ halos. The shaded region above the upper 
line implies less than 1 halo within a Hubble radius which is meaningless. The 
$3\;10^4$ and $10^{10}$ halo lines roughly correspond to one halo per 
supercluster (every 4~$10^6$~Mpc$^3$) and to one per galaxy (every 500~Mpc$^3$)
respectively. Note that this is a maximum number of halos, for instance below 
the lower curve models can still have clusters around all galaxies, as long as 
the population of neutrinos in between the halos is non zero. 

Notice that protons are attenuated in the CMB in an energy loss distance of 
about 50 Mpc. This means that for $D > 50$ Mpc the region of the halo outside 
a sphere of this radius centered around us can be ignored for the production 
of the local UHECR spectrum. Halo sizes exceeding 50 Mpc should be considered 
in these plots as having an effective size of 50~Mpc. 

The approach is very conservative. Eq.(\ref{eneutrino}) neglects fractions of 
the $Z^0$ production energy which goes into particles that cannot be UHECR. 
The $Z^0$ decay will produce a particle flux following a typical fragmentation 
spectrum and the decay of the unstable particles will add low energy particles 
which cannot contribute to the UHECR. Neither can neutrinos from $Z^0$ and pion
decays nor that part of the high energy particles that are degraded by the 
showering developed in the intergalactic medium. The proton energy flux in 
Eq.(\ref{eproton}) depends on the limits of the integration. It has been 
conservatively estimated by setting them close to the UHE part of the CR 
spectrum. As the observed cosmic ray flux spectral index at these energies is 
about 2.5, the lower integration limit gives the dominant contribution to 
the integral. The upper limit is not so relevant and it is in any case bounded 
by the neutrino resonance energy which in turn depends on the neutrino mass.
For harder spectral indices closer to 2 one gets a similar result using an 
upper limit of order $E_{max}=100~E_{min}$ as in Ref.~\cite{Waxman}. 

We have also done an analytical calculation of the proton and photon 
fluxes originated from the neutrino--antineutrino annihilation again 
using ${\cal P}_I(E_{\nu})$ obtained from  Eq.~(\ref{localopacity}). 
The flux of protons is given by:
\begin{equation}
\phi_p(E_p)= \int_0^\infty \phi_\nu(E_\nu) \frac{d \sigma(\nu \bar \nu 
\rightarrow p)}{d E_p} \bar X,
\end{equation}
where $d\sigma/dE_p$ is the cross section for the $\nu \bar \nu$ to produce a 
proton of energy $E_p$ and $\bar X$ is the column depth of neutrinos in the 
halo. The cross section $d\sigma/dE_p$ can be written as the convolution:
\begin{equation}
\frac{d \sigma(\nu \bar \nu \rightarrow p)}{dz}= \int_z^1 dy \frac{d
\sigma(\nu \bar \nu \rightarrow q)}{d y} f(z/y), 
\end{equation}
where $z$ is the fraction of the energy taken by the proton, $f(x)$ is the 
fragmentation function of a quark into an hadron, and $d \sigma(\nu \bar \nu 
\rightarrow q)/dy$ is the inclusive cross section for quark production from 
$\nu \bar \nu$ interactions. At the $Z^0$ resonance the non-resonant channels 
can be neglected at this level of precision. We use Hill's fragmentation 
function \cite{Hill}, $f(x)= N \; 15/16 x^{-1.5} (1-x)^2$ with $N= 0.03$ for 
baryons. For calculating the secondary UHE photon spectrum similar expressions 
apply with $N = 0.32$ and an additional integral over $\pi^0$ decay. These 
integrals are evaluated numerically. We normalize the photon plus proton 
spectrum to the observed cosmic ray flux at $E>5 \; 10^{19}$ eV. We neglect 
the interaction of the high energy particles produced with the IR and CMBR 
which would increase the neutrino flux needed. 

If we again apply the Fly's Eye limit, parameter space becomes more restricted 
leaving less room for the conjecture as can be seen in fig.~\ref{fig3}. 
No model with $\gamma > 2$ is allowed in agreement with 
Ref.~\cite{Waxman} but there is still room for harder spectral indices. 
The natural assumption of one halo per galaxy forcing small halo sizes and 
bounding the possible density enhancements is ruled out for any injection 
spectrum. This is unfortunate since a clear experimental signature of 
relatively small halo sizes would be a cosmic ray anisotropy due to the 
asymmetric position of the solar system within the halo, given by the ratio 
of our position and the galactic halo radius $D$. Assuming $10\%$ sensitivity 
to anisotropy a future experiment such as the Auger Observatories \cite{Auger} 
could test models with halos of order 100 kpc. The 
picture is however not complete since neither Pauli blocking nor total mass 
constraints have been included. These must be related to mass density bounds 
that exist on different scales. 

The two case studies in Ref.~\cite{Yoshida} use a halo size of 5 Mpc and 
density enhancements of 300 and 1000 ($\xi \sim 2.5$ and $\xi \sim 3$). 
As stated in Ref.~\cite{Yoshida} they are not in conflict with horizontal 
air shower data; this is because of the very hard spectral indices 
used, $\gamma \sim 1.2$. It 
is interesting to notice, however, that the maximum number of halos 
corresponding to these models is $\sim 7 \; 10^5$ and $\sim 2 \; 10^5$ 
respectively, implying that most galaxies can not 
have an associated neutrino halo, yet the size of these halos is of 
order the average inter galactic distance. 

The total mass in 
neutrinos of such enhanced density halos is, on the other hand:
\begin{equation}
M_\nu= \frac{4}{3} \pi \; m_\nu \; n_\nu \; D^3 \; = 1.2 \; 10^{10} \; D^3 
10^\xi \; \left( \frac{m_\nu}{1 \mbox{eV}} \right) \; M_{\odot}
\end{equation}
Assuming a 1 eV neutrino mass this gives at least 
$M_\nu = 4.5 \; 10^{14} M_{\odot}$ for the most favourable case with 
density enhancement 300. 
Although this mass is suggestive of supercluster scales, its size fits 
closer to the Local Group. Unfortunately the case studies imply a total 
mass in neutrinos that clearly exceeds the dynamic mass measurements 
associated to the Local Group. For $\Omega=1$ the mass within a sphere 
of radius $R$ given by $H_0 R= 390$~km~s$^{-1}$ is determined to be 
$M_G=5.7~10^{12} M_{\odot}$ \cite{Peebles}. The mass increases by 
about $10 \%$ for $\Omega=0.1$. Although the horizontal shower data 
and the halo mass constrains can be both met by a model neutrino mass 
($m_{\nu}< 0.1~$eV) and spectral index $\gamma < 1.2$, Pauli blocking 
arguments would imply orbital velocities exceeding $1000~$km~s$^{-1}$ 
implying that a constant density model is inconsistent. 

The only other possibility that is left for these models is an even larger 
halo size. A halo size of $\sim 50$~Mpc, a neutrino 
spectral index of $\gamma \sim 1.2$ and a neutrino mass of order  
$0.1~$eV seems viable. This case corresponds to a halo mass of order 
$10^{15} M_{\odot}$ and orbiting velocities in the 1000~km$^{-1}$ range.  Increasing the halo size helps mostly by removing the 
strong constrain on the total mass. 

In summary Horizontal Shower limits, 
mass constraints and Pauli blocking mechanisms leave very little room 
for UHE neutrinos to explain the origin of the UHECR. Only very large 
halos, relatively low neutrino masses and unusually flat spectral indices 
are marginally allowed. Up to now we have assumed that there is absolute 
lepton number symmetry and in such cases 
the density parameter for the neutrinos is fixed by the neutrino mass: 
\begin{equation}
\Omega_{\nu}= {1 \over h^2} {m_{\nu} \over 93~\mbox{eV}} 
\end{equation}
It is however remarkable that if there is lepton number asymmetry, as recently 
suggested in Ref.\cite{Gelmini}, density enhancement comes rather naturally 
and is distributed uniformly over the whole Universe. 
Assuming that neutrinos are degenerate, $\Omega_\nu \sim 0.01$ and $m_\nu\sim 
0.07$ eV, they obtain an enhancement factor of $\sim 30 $. Our 
analysis has been repeated considering the interactions within a sphere 
of $\sim 50~$Mpc, revealing that the model is completely consistent 
with the Fly's Eye limit on neutrino fluxes, provided the spectral 
index of the UHE neutrinos satisfies $\gamma \lesssim 1.8$. 

In any case there is no experimental signature provided by anisotropy. 
In the end however it is fortunate that these models can be further 
tested by experiment. A promising signature 
lies in the identification of photons as a significant component of the 
UHECR. This issue can be addressed by future experiments such as the 
Auger Observatories \cite{Auger}. 
Most importantly the fact that horizontal showers provide such a 
strict bound on these models also implies that future neutrino 
experiments, having much larger acceptance for neutrinos than Fly's Eye, 
should be able to detect the postulated UHE neutrino fluxes. Here the 
Auger Observatories may also play a role together with other high
energy neutrino detectors in construction or planning stages. 


We would like to thank Concha~Gonzalez-Garcia for many discussions 
after reading the manuscript and Edward Baltz and Joseph Silk for 
discussions on mass bounds and Graciela Gelmini, Alexander Kusenko, 
Ken D. Olum, G\"unter Sigl, and Alex Vilenkin for helpful conversations. 
This work was supported in part by Xunta de Galicia (XUGA-20602B98), by 
CICYT (AEN99-0589-C02-02), and the Fundaci\'on Pedro Barrie de la Maza.

\end{document}